\documentclass[aps,twocolumn,superscriptaddress]{revtex4-2}
\usepackage{upgreek}
\usepackage{graphicx}
\usepackage{bm}
\usepackage{amsmath, amsthm, amssymb}
\usepackage{latexsym}
\usepackage{dcolumn}
\usepackage{color}
\usepackage{mathtools}

\DeclarePairedDelimiterX\braket[2]{\langle}{\rangle}{#1 \delimsize\vert #2}

\begin{document}

%\preprint{}
\title{Multiplexed sensing of magnetic field and temperature in real time using a nitrogen-vacancy spin ensemble in diamond}

\author{Jeong Hyun Shim}
\email[]{jhshim@kriss.re.kr}
\affiliation{Korean Research Institute of Standards and Science, Daejeon 34113, Republic of Korea}
\affiliation{Department of Medical Physics, University of Science and Technology, Daejeon 34113, Republic of Korea}
\author{Seong-Joo Lee}
\affiliation{Korean Research Institute of Standards and Science, Daejeon 34113, Republic of Korea}
\author{Santosh Ghimire}
\affiliation{Korean Research Institute of Standards and Science, Daejeon 34113, Republic of Korea}
\author{Ju Il Hwang}
\affiliation{Department of Physics, Hanyang University, Seoul 04763, Republic of Korea}
\author{Kwang-Geol Lee}
\affiliation{Department of Physics, Hanyang University, Seoul 04763, Republic of Korea}
\author{Kiwoong Kim}
\affiliation{Department of Physics, Chungbuk National University, Cheongju 28644, Republic of Korea}
\author{Matthew J. Turner}
\affiliation{Department of Physics, University of Maryland, College Park, Maryland, 20742, USA}
\affiliation{Department of Electrical Engineering and Computer Science, University of Maryland, College Park, Maryland, 20742, USA}
\affiliation{Quantum Technology Center, University of Maryland, College Park, Maryland, 20742, USA}
\author{Connor A. Hart}
\affiliation{Department of Physics, University of Maryland, College Park, Maryland, 20742, USA}
\affiliation{Department of Electrical Engineering and Computer Science, University of Maryland, College Park, Maryland, 20742, USA}
\affiliation{Quantum Technology Center, University of Maryland, College Park, Maryland, 20742, USA}
\author{Ronald L. Walsworth}
\affiliation{Department of Physics, University of Maryland, College Park, Maryland, 20742, USA}
\affiliation{Department of Electrical Engineering and Computer Science, University of Maryland, College Park, Maryland, 20742, USA}
\affiliation{Quantum Technology Center, University of Maryland, College Park, Maryland, 20742, USA}
\author{Sangwon Oh}
\email[]{sangwon.oh@kriss.re.kr}
\affiliation{Korean Research Institute of Standards and Science, Daejeon 34113, Republic of Korea}


\date{\today}

\begin{abstract}
Nitrogen-Vacancy (NV) spin in diamond is a versatile quantum sensor, being able to measure physical quantities such as magnetic field, electric field, temperature, and pressure. In the present work, we demonstrate a multiplexed sensing of magnetic field and temperature. The dual frequency driving technique we employ here is based on frequency-division multiplexing, which enables sensing both measurables in real time.
 The pair of NV resonance frequencies for dual frequency driving must be selected to avoid coherent population trapping of NV spin states. With an enhanced optical collection efficiency higher than 50 $\%$ and a type 1b diamond crystal with natural abundance $^{13}$C spins, we achieve sensitivities of about 70 pT/$\sqrt{\mathrm{Hz}}$ and 25 $\mu$K/$\sqrt{\mathrm{Hz}}$ simultaneously. A high isolation factor of 34 dB in NV thermometry signal against magnetic field was obtained, and we provide a theoretical description for the isolation factor. This work paves the way for extending the application of NV quantum diamond sensors into more demanding conditions.
\end{abstract}

%\keywords{}

\maketitle

\section{Introduction}
%\label{Introduction}

An ensemble of negatively-charged Nitrogen-Vacancy (NV) spins in diamond can provide a broadband magnetic field sensitivity in the picotesla range, depending on instrument details \cite{Barry2016, Ahmadi2017, Chatzidrosos2017, Schloss2018, Zheng2019, Fescenko2020, Webb2021, Barry2020}. For such quantum magnetometers, precise measurement of the NV spin resonance frequencies is essential. The energy levels of NV spin states, however, depend on other physical quantities \cite{Doherty2012}  as well, such as temperature\cite{Acosta2010}, electric field \cite{Dolde2011}, strain \cite{Trusheim2016}, and pressure \cite{Doherty2014}. Thus, when exploited as a magnetometer, NV spins need to be isolated from variations of other quantities to read the correct value of magnetic field. Under ambient conditions, where the influences of electric field and pressure are reasonably suppressed, thermal drift may be a particular challenge. In principle, the more sensitive a diamond magnetometer, the stronger it is affected by thermal variation. In other words, a sensitive diamond magnetometer can function as a sensitive thermometer too. There have been early works demonstrating pulse sequences on NV spins in order to operate selectively as either a magnetometer or a thermometer. For example, double quantum sequences allow NV spins to accumulate the phase induced only by magnetic fields \cite{Mamin2014}, while with alternative sequences, such as D-Ramsey, the phase is sensitive only to temperature \cite{Neumann2013}.  For multiplexing these measurables, continuous-wave (cw) operation has to date relied on time-division, i.e., alternating quantities projected by optical readout, leading to sequential monitoring of magnetic field and temperature with a specific time step \cite{Clevenson2015, Clevenson2018, Hatano2021}. %In time-division multiplexing or sequential measurement, a sensor is temporarily insensitive to other measurables, which will deteriorate the sensitivity and bandwidth.
Frequency-division multiplexing enables implementing a real-time recording of multiple quantities because multiple frequency components can coexist in a continuously varying signal \cite{Wojciechowski2018, Schloss2018}. In this paper, we demonstrate that, with dual frequency driving \cite{Fescenko2020}, magnetic field and temperature can each be measured concurrently in real time for an NV ensemble. Improved optical design in the NV fluorescence detection enabled us to achieve a magnetic field sensitivity of 70 pT/$\sqrt{\mathrm{Hz}}$ and a temperature sensitivity of 25 $\mu$K/$\sqrt{\mathrm{Hz}}$ simultaneously, and with an isolation factor of about 34 dB, using a high-pressure high-temperature (HPHT) grown diamond crystal of natural $^{13}$C abundance and NV density of about 0.5 ppm (see further information below).
\begin{figure*}[t]
\centering
\includegraphics[origin=c,clip=true,width=0.95\textwidth]{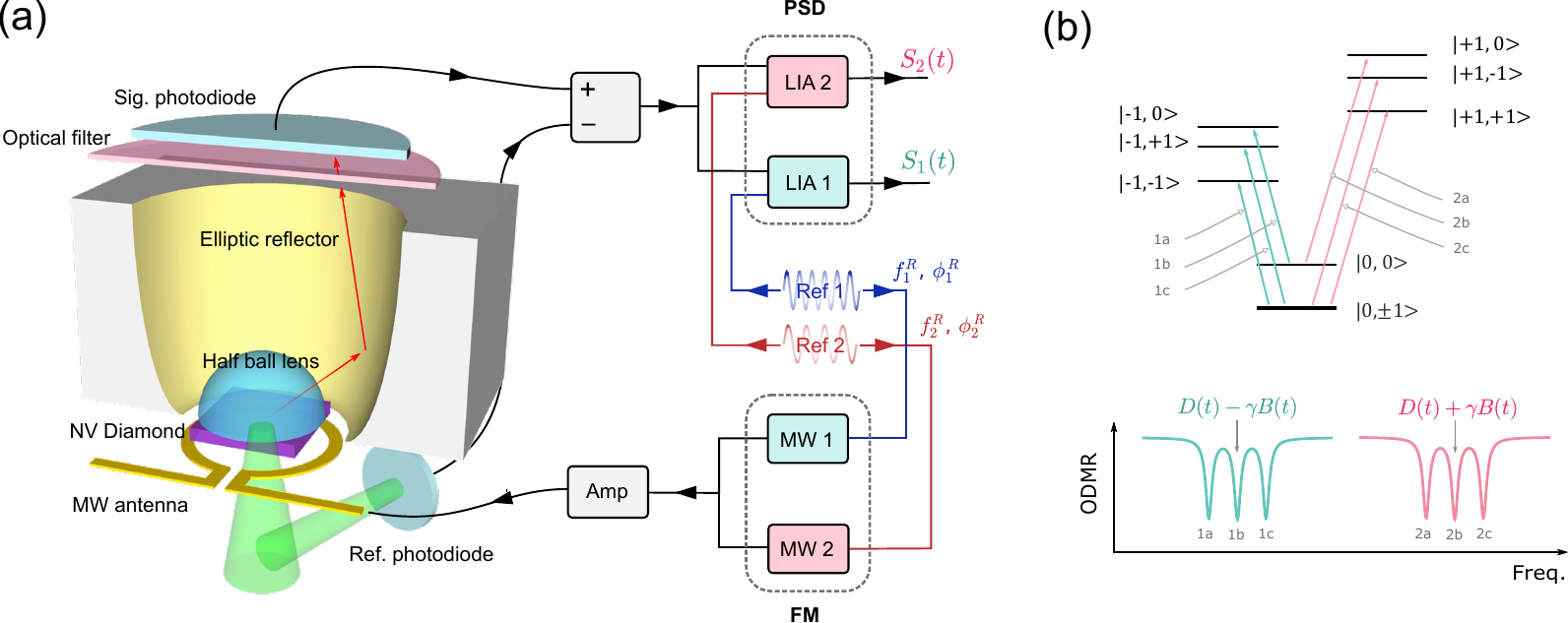}
\caption{(a) Configuration of NV diamond sensor with microwave driving and lock-in detection for dual frequency driving. A half ball lens on the top of the NV diamond sample in combination with an elliptic reflector guides the fluorescence from NV spins to the signal photodiode. The reference photodiode monitors the green (532nm) pump beam and is used to suppress laser intensity fluctuations by subtracting the output of the reference photodiode from the signal photodiode using a balanced circuit. Reference 1 and 2 are fed into microwave sources (MW1, MW2) and lock-in amplifiers (LIA1, LIA2) for frequency modulation (FM) and phase-sensitive detection (PSD), respectively. The frequencies ($f_1^R$, $f_2^R$) and phases ($\phi_1^R$, $\phi_2^R$) of the reference signals (Ref1, Ref2) for FM are controllable parameters for dual frequency driving (described in the text). (b) (up) Energy diagram of NV electronic spin coupled with $^{14}$N nuclear spin, indicating six transitions. (down) ODMR spectrum indicates the six transition frequencies. }
\label{fig:1}
\end{figure*}

\section{Results}
\subsection{Background}
The shot-noise limited sensitivity $\eta_B$ of a cw diamond magnetometer is governed by intrinsic properties of NV spins and extrinsic configurations of optical detection, and can be expressed as\cite{SM, Hobbs2009}
\begin{equation}
		\eta_B = \frac{\Delta f}{\frac{\gamma}{2 \pi} C} \sqrt{\frac{2 e}{i_{ph}}}.
\end{equation}
$\gamma$ and $e$ are fundamental constants as gyromagnetic ratio and electron charge, respectively. $\Delta f$ and $C$ are linewidth and contrast in cw optically-detected magnetic resonance (ODMR) obtained via lock-in detection. $\frac{\Delta f}{C}$ is associated with the inverse of zero-crossing slope ($\alpha$), converting the voltage-noise, induced by the shot noise of the photo-induced reverse current ($i_{ph}$) out of a photodiode, to magnetic field noise (See Supplementary Information S1). Given $\eta_B$, the temperature sensitivity $\eta_T$ is anticipated as,
\begin{equation}
		\eta_T = \frac{\gamma }{2 \pi \kappa} \eta_B,
\end{equation}
in which $\kappa$ = $|\frac{\partial D}{\partial T}|$ = 74.2 kHz/K (D: zero-field splitting) at room temperature \cite{Acosta2010}. Typically, obtainable values for an optimized NV ensemble\cite{Barry2020} ($\Delta f$= 1 MHz, $C$ = 1 \%, $i_{ph}$ = 10 mA) lead to the sensitivities as $\eta_B$= 20 pT/$\sqrt{\mathrm{Hz}}$ and $\eta_T$ = 7.6 $\mu$K/$\sqrt{\mathrm{Hz}}$, supporting the above statement that NV diamond can serve as a highly-sensitive thermometer too. Although $\Delta f$ and $C$ are dependent on pumping laser and microwave driving, such as number of resonances and power (P$_{\mathrm{MW}}$) and etc., their limits are also strongly influenced by the quality of a diamond, in terms of the amount of the sources of line-broadening and charge-trapping \cite{Barry2020, Edmonds2021}.
\subsection{NV diamond sensor}
In the present work, we used a type 1b HPHT diamond crystal with natural $^{13}$C abundance and initial Nitrogen (P$_1$ center) density of approximately 30 ppm (see Methods in appendix). NV centers were created by electron irradiation and annealing at high temperature, sequentially. The residual concentrations of Nitrogen (N$_{\mathrm{S}}^0$) and NV- center in NV diamond were estimated 1.3 ppm and 0.5 ppm, respectively. In order to increase $i_{ph}$ in Eq. (1), collection optics are designed to increase the collection of the fluorescence signal from NV centers and to deliver it to the signal photodiode as depicted in Fig. 1(a). A half-ball lens made of a high-refractive index material (n $\cong$ 2.0, S-LAH79) was used to alleviate the light trapping inside the high index diamond region (n = 2.4). Numerical calculation shows that the solid-angle of the photon escape cone is nearly five times as high as that without the half-ball lens, and the total photon collection efficiency is approximately 56 \% (See Supplementary Information S2).

\subsection{Dual frequency driving}
The Hamiltonian for the NV electronic spin ($S=1$) coupled to an $^{14}$N nuclear spin ($I=1$) can be expressed as $\mathcal{H} = DS_z^2+ \gamma B S_z + A_{zz} S_z I_z - PI_z^2$ with zero-field splitting $D$, magnetic field $B$ along the NV spin orientation ($B_x, B_y = 0$), $^{14}$N quadrupole splitting $P$, and the hyperfine coupling $A_{zz}$ \cite{Levine2019}. With a static magnetic field aligned along [111] orientation of the diamond crystal, the NV spins parallel to the field reveals six transitions in the ODMR spectrum, as shown in Fig. 1(b). The center ($f_-$) of the lower frequency triplet corresponds to $D(t) - \gamma B(t)$, while that ($f_+$) of the higher frequency to $D(t) + \gamma B(t)$, assuming that zero-field splitting and magnetic field are time varying parameters as $D(t) = D_0 + \Delta D(t)$ and $B(t)=B_0 + \Delta B(t)$; $D_0$ and $B_0$ are time-invariants.

\begin{figure}[t]
\centering
\includegraphics[width=\columnwidth]{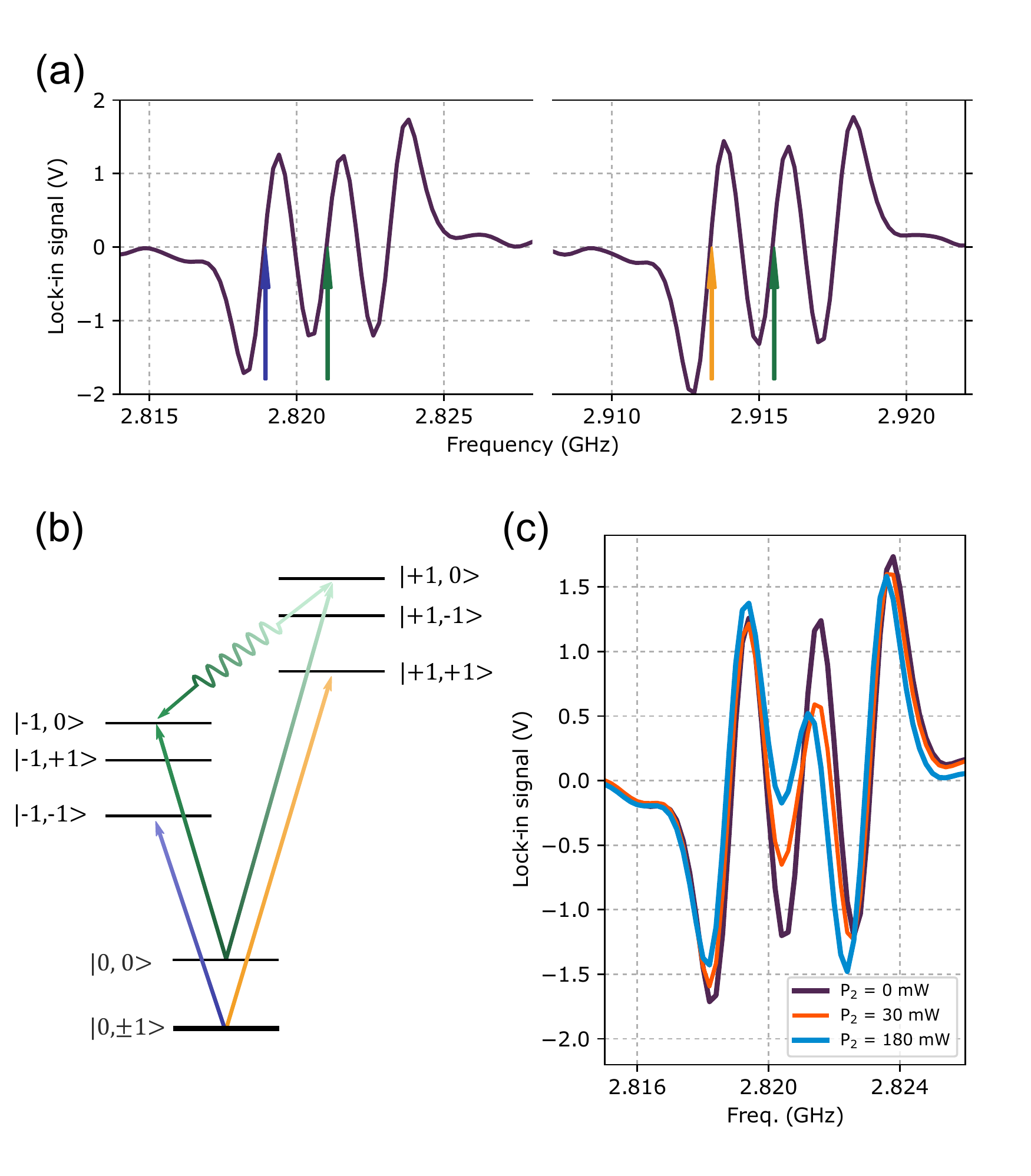}
\caption{(a) ODMR spectrum obtained by phase sensitive detection using LIA. The colored arrows indicate the frequencies of the transitions represented in (b) with the same colors. (b) The transition pair (green) have Rabi frequencies of $\Omega_+$ and $\Omega_-$ and form a $V$ type level configuration, which leads to coherent bright $|B\rangle$ and dark $|D\rangle$ states; in contrast, the other pair (blue, yellow) does not. (c) Coherent population trapping (CPT) involving $|B\rangle$ and $|D\rangle$  supresses the ODMR contrast. The suppression becomes higher as the power of MW2 (P$_2$) increases. }
\label{fig:2}
\end{figure}
A dual frequency driving scheme lies at the heart of multiplexed sensing of magnetic field and temperature. We apply two microwaves (MW1, MW2) driving the two transitions centered at $f_-$  and $f_+$  at the same time. Both are frequency-modulated according to reference signals (Ref1, Ref2), shown in Fig. 1(a), which are also fed into two lock-in amplifiers (LIA1, LIA2) independently. The frequencies ($f_1^R$, $f_2^R$) and phases ($\phi_1^R$, $\phi_2^R$) of Ref1 and Ref2 can be adjusted depending on which quantity needs to be measured via NV diamond. For the multiplexed sensing, setting $f_1^R$ unequal to $f_2^R$ and the difference $|f_1^R - f_2^R |$ larger than the bandwidth of LIA is necessary. Then, the outputs of LIA1 ($S_1$) and LIA2 ($S_2$) will be given as $S_1 (t)= \alpha [ \Delta D(t) -\gamma \Delta B(t)]$, $S_2 (t)=\alpha [ \Delta D(t) + \gamma \Delta B(t) ]$, respectively, where $\alpha$ is the slope of the LIA output. By adding and subtracting $S_1 (t)$ and $S_2 (t)$, we can obtain both the magnetic field $\Delta B(t)$ and thermal $\Delta D(t)$ variations in real time. If multiplexing is not necessary, the more common configuration is to adopt only a single LIA by setting $f_1^R$ equal to $f_2^R$ \cite{Wojciechowski2018, Fescenko2020}. With $\phi_1^R$ - $\phi_2^R$ = $\pi$, the output of LIA becomes $S_B (t) = 2 \alpha \Delta B(t)$; thus the NV diamond sensor functions as a magnetometer with a doubled contrast or zero-crossing slope. Conversely, the NV diamond sensor acts a thermometer when $f_1^R = f_2^R$  but $\phi_1^R = \phi_2^R$, since $S_T (t) = 2\alpha \Delta D(t)$.

\subsection{The effect of coherent population trapping}
There are 9 possible combinations of hyperfine transitions that could be driven. Ideally the pair of the middle hyperfine lines as indicated by green arrows in Fig. 2 (a) can be used to make the lineshape symmetric. We found, however, that such combination should be excluded due to diminished ODMR contrast caused by coherent population trapping (CPT) \cite{Kehayias2014, Jamonneau2016}. The need to exclude transitions impacted by CPT effects necessitates further analysis on the extracted magnetometry and thermometry signals to mitigate nonlinear effects \cite{SM}. The detailed analysis of CPT on NV spins with cw microwave and laser is out of the scope of the present paper, therefore we give a brief phenomenological explanation as follows. In Fig. 2 (b), the two transitions indicated by green arrows can form a $V$ type level configuration, because the two excited states $| -1,0\rangle$ and $| +1,0\rangle$ possess the same $^{14}$N state $m_I = 0$. The two microwaves, then, coherently interact and render a bright $| B\rangle$ and a dark $| D \rangle$ state on NV spins, which are linear superposition of $| +1\rangle$ and $| -1\rangle$ \cite{Jamonneau2016, shim2013} as,
\begin{eqnarray}
|B\rangle &=& \frac{1}{\sqrt{ \Omega_+^2 + \Omega_-^2}} [ \Omega_- |-1 \rangle + \Omega_+ |+1\rangle ] \nonumber\\
|D\rangle &=& \frac{1}{\sqrt{ \Omega_+^2 + \Omega_-^2}} [ \Omega_+ |-1 \rangle - \Omega_- |+1\rangle ]
\end{eqnarray}
When there is an irreversible decay from $|B\rangle$ to $|D\rangle$, the population of the excited state is trapped to the dark state. This trapping process, however, competes with the decoherence in the excited states and the repolarization to $| 0\rangle$, caused by the laser pumping. As $\Omega_-$ increases, the dark state becomes closer to $|-1\rangle$ since $|\langle -1 | D \rangle|^2 = \frac{\Omega_+^2}{ \Omega_+^2 + \Omega_-^2}$. Then, presumably the trapping is less affected by the laser-induced decoherence. This explanation is further supported by increased suppression of ODMR contrast with a higher mw power at $f_+$  as shown in Fig. 2(c). Thus, one needs to avoid CPT in order to take advantage of doubled contrast when using dual frequency driving. For the results shown in Fig. 3 and Fig. 4, we adopted a mw pair indicated by blue and yellow arrows in Fig. 2 (a) and (b), which eludes a $V$-type configuration.

\begin{figure}[t]
\centering
\includegraphics[width=\columnwidth]{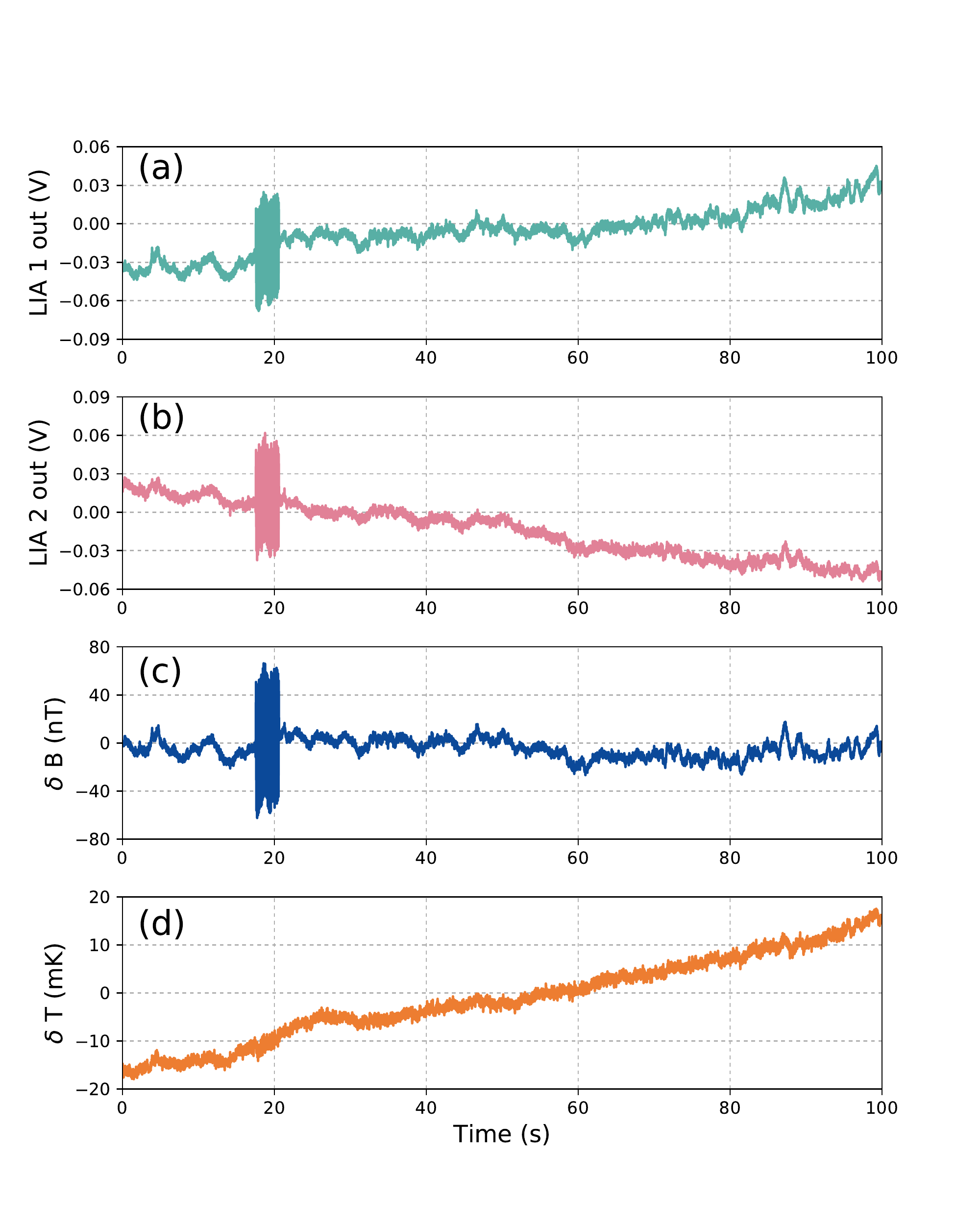}
\caption{(a), (b) Time traces of the outputs of LIA1 and LIA2 in Fig. 1(a). The summed and subtracted signals are shown in (c) and (d), respectively. A magnetic field of 38.9 nT (RMS) at 10 Hz was applied for 3 s. The vertical axes of (c) and (d) are scaled according to the calibration shown in Fig. 4 (b) and (c). The applied field is visible only in the summed signal, while the subtracted signal senses temperature variation. }
\label{fig:3}
\end{figure}
\subsection{Multiplexed sensing of magnetic field and temperature}
Figure 3(a) and (b) are the real-time traces of the outputs of LIA1 and LIA2, respectively. The digitized signals are, then, subtracted (Fig. 3(c)) and summed (Fig. 3(d)) in software. The subtracted signal manifests rapid variations, while the summed signal exhibits a relatively smoother and gradual increment. To verify that the subtracted signal measures magnetic field variation only, we applied a magnetic field modulation of amplitude 38.9 nT (RMS) at 10 Hz for 3 s and found that the test field signal appears only on the subtracted time trace. The noticeable differences between Fig. 3(c) and (d) can be explained by a fluctuation of ambient magnetic field at low frequency, which appears in the noise spectrum in Fig. 4(a). To confirm that the summed signal interrogates temperature, we measured the temperature of the NV diamond sample with a thermal sensor (PT100) mounted nearly 5 mm away from the NV diamond on the sapphire plate \cite{SM} and correlated it with the summed signal as shown in Fig. 4(c). A linear correlation was obtained, with a slope that is used for the calibration of the vertical scale of the summed signal (Fig. 3(d)). Similarly, Fig. 4 (b) shows the signal obtained with a known magnetic field of 1 $\mu$T (RMS), applied for the calibration of the vertical scale of the subtracted signal (Fig. 3(c)).

\subsection{Sensitivity}
For the estimation of the magnetic field and temperature sensitivity, the output of LIA1 was fed into a FFT spectrum analyzer directly in order to bypass noise contributions during digitizing. Here, the reference frequency $f_1^R$ was set equal to $f_2^R$ for operating NV diamond as either a magnetometer or a thermometer respectively. The noise spectra in Fig. 4 (a) contains the responses to a test magnetic field of 1 $\mu$T on top of the observed noise floor. The noise spectrum of the NV diamond magnetometry signal includes the main test signal peak at 10 Hz and higher harmonics indicated by gray arrows. The power line (60 Hz) noise is marked by the red arrow. The 1/f noise below about 4 Hz is also suppressed in the NV diamond thermometry signal. This indicates that the ambient environment contains pronounced low-frequency magnetic field fluctuations.  The noise floor of the NV diamond thermometer is lower than that of the magnetometer for frequencies up to 100 Hz. This is also attributed to the further suppression of ambient magnetic noise. When all the microwaves are turned off, the noise level decreases by about 3 dB. With the intensity of the test field and the slope in Fig. 4 (c), the noise floors are converted into the sensitivities of magnetometer and thermometer as 70 pT/$\sqrt{\mathrm{Hz}}$ and 25 $\mu$K/$\sqrt{\mathrm{Hz}}$.

\subsection{Isolation factor}
In Fig. 4 (a), the main peak at 10 Hz of the NV thermometry signal is significantly reduced by 34 dB, which we define as isolation factor. Such high isolation factor indicates temperature measurement is well isolated from the magnetic field. We performed theoretical and numerical analyses, in order to investigate the limitations of the isolation factor and the presence of higher harmonics\cite{SM}.  We note three dominant effects: (1) off-axis magnetic field contributions in the Hamiltonian, (2) nonlinear effects due to the Lorentzian ODMR lineshape, and (3) the error in balancing the zero-crossing slopes. Our numerical analysis revealed that the nonlinear effect contributes only to the even harmonics (20 Hz, 40 Hz, etc.), while the balancing error can significantly reduce the isolation factor (10 Hz). We, thus, minimized the main peak intensity at 10 Hz in the NV thermometry signal by controlling parameters (see further information below). When the balancing error is negligible, 2nd order terms in Hamiltonian, involving the off-axis components $B_x$ and $B_y$, should be considered to explain the isolation factor. We assume that the time-varying field, like the oscillating test field at 10 Hz, also has off-axis components $\Delta B_x$ ($\Delta B_y = 0$), but $\Delta B_x  < B_x$. Then, the variation of NV resonance frequency $\Delta f_+$ and $\Delta f_-$ can be expressed as \cite{Turner2020}
\begin{equation}
\Delta f_{\pm} = (\frac{\partial{f}}{\partial{T}})\Delta T \pm \frac{\gamma}{2 \pi} \Delta B_z + \frac{3 }{D} (\frac{\gamma}{2 \pi})^2 B_x \Delta B_x,
\end{equation}
in which $\Delta T$ and $\Delta B_z$ represent the changes of temperature and magnetic field along NV axis, respectively. The ${\Delta B_x}^2$ term is omitted since it's smaller than the $B_x \Delta B_x$ term. Equation (4)  indicates that off-axis magnetic field component can couple into the thermometry signal $S_T (t)$ through a 2nd order term, since $S_T (t) = 2\alpha [(\frac{\partial f}{\partial T}) \Delta T +  \frac{3 }{D} (\frac{\gamma}{2 \pi})^2  B_x \Delta B_x]$ . The isolation factor $\xi$ can be obtained as
\begin{equation}
\xi = \frac{2 \pi D \Delta B_z}{3 \gamma B_x \Delta B_x},
\end{equation}
which reveals that reducing off-axis components is crucial in enhancing the isolation factor. According to the experimental configuration we adopted, $\Delta B_z$ = 1 $\mu$T (RMS) and $\Delta B_x$ = $\sqrt{2}$ $\mu$T (RMS) \cite{SM}. From the experimentally obtained value of $\xi$= 2511 (34 dB), the off-axis field $B_x$ was estimated to be 8.37 $\mu$T, corresponding to a misalignment angle of 0.3$^{\circ}$.

\begin{figure}[t]
\centering
\includegraphics[width=\columnwidth]{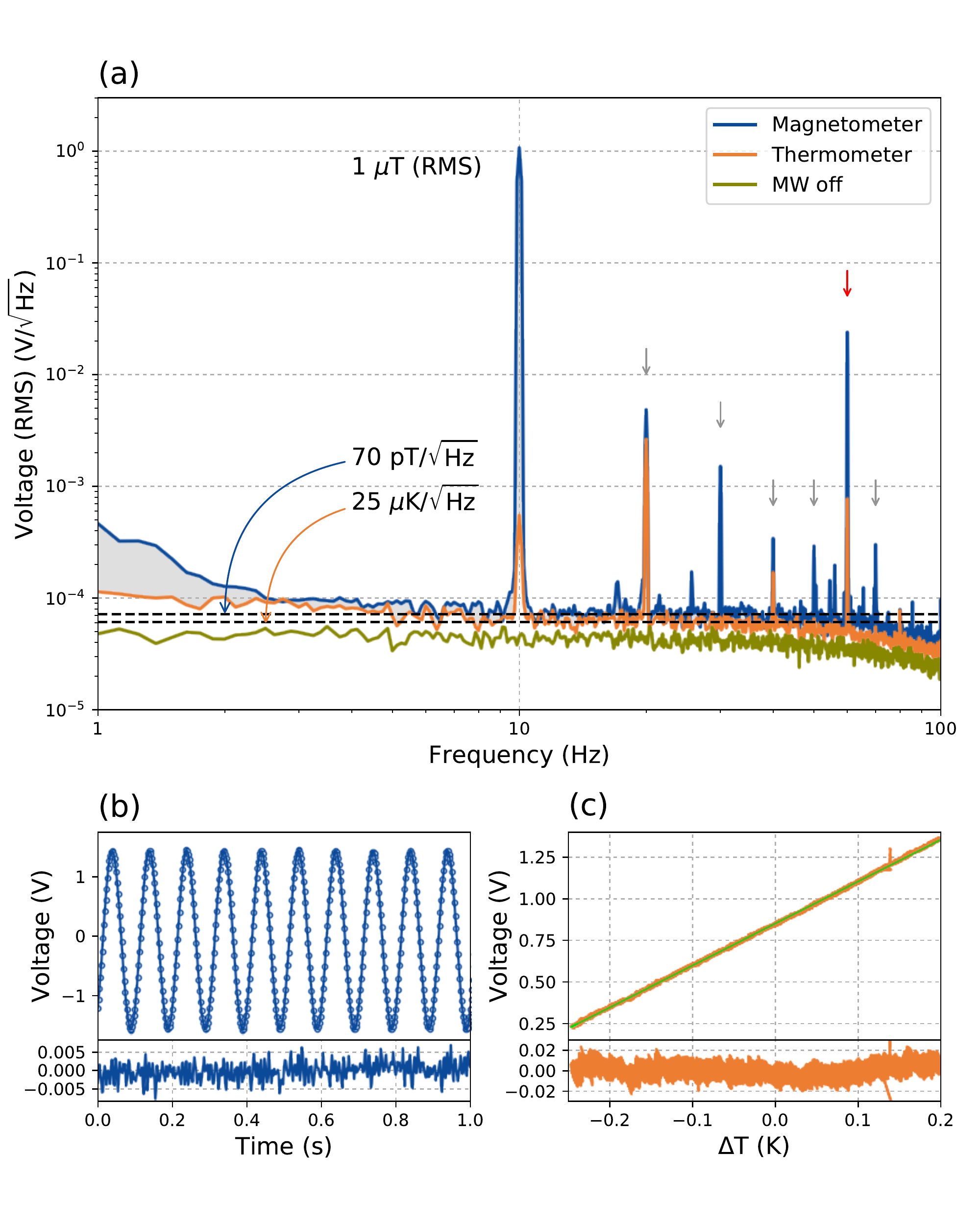}
\caption{(a) Noise spectra of NV diamond magnetometry and thermometry signals. The system noise is characterized without microwaves (MW off). A test field of 1 $\mu$T (RMS) at 10 Hz is applied for estimating the isolation factor and for calibration. When the NV sensor functions as a thermometer, the signal due to the test field is reduced by about 34 dB. The peaks indicated by gray arrows are higher harmonics of the test field and the red arrow indicates power line noise. The gray area below about 4 Hz represents 1/f magnetic field noise. (b) The amplitude of the test field response is obtained from numerical fitting and is used for calibration of the NV magnetometer. (c) Similarly, the temperature of the NV diamond sample is measured with a PT100 sensor and recorded for over 2 hours in addition to the signal of the NV thermometer. As expected, a linear response is obtained, of which the slope is used for the calibration of NV thermometer. After calibration, the sensitivities of the NV magnetometer and NV thermometer are determined to be about 70 pT/$\sqrt{\mathrm{Hz}}$ and 25 $\mu$K/$\sqrt{\mathrm{Hz}}$, respectively. }
\label{fig:4}
\end{figure}
\section{Discussion}
In the spectrum of the NV thermometer in Fig. 4(a), peaks are observed mainly at even harmonic frequencies, except for the main peak (10 Hz) for which the intensity is determined by the isolation factor. We conducted numerical simulations with an assumption that NV spin resonances, measured by ODMR, have a Lorentzian lineshape; and without including 2nd order terms in Eq.(4) to isolate the effects of the nonlinear lineshape. Our numerical simulation attributes those higher harmonic peaks to the nonlinearity in the zero-crossing slope of the derivative of Lorentzian lines. According to the simulation, dual frequency driving for the NV thermometer effectively eliminates odd harmonic peaks, leaving only even harmonics. This simulation may explain the observed NV thermometer’s noise spectrum\cite{SM}. For the NV magnetometry signal, the nonlinear lineshape simulation shows that only odd harmonic peaks can survive the measurement protocol, which is inconsistent with the 20 Hz and other even harmonic magnetometry peaks in the spectrum in Fig. 4(a). We used further simulations to attribute these additional even harmonics of the magnetometry signal to imperfections in matching the two zero-crossing slopes of $f_+$ and $f_-$ and imperfections in the drive frequency. For ideal operation of dual frequency driving, the slopes and responses of $f_+$ and $f_-$ should be matched as much as possible. Since the zero-crossing slopes depend significantly on MW power and polarization, achieving identical slopes is challenging in practice for an NV ensemble and needs to be tuned. According to the simulations, a mismatch of 1 to 5 \% is sufficient to produce a 2nd harmonic peak in the NV magnetometry signal similar to that observed in the experiment. For the case of the NV thermometry signal, we minimized the peak intensity at 10 Hz by tuning the phase $\phi_1^R$, $\phi_2^R$ of the reference signals, which optimizes balancing. The effect of nonlinearities can be reduced through a fast feedback on the MW center frequency to stay in the linear regime as much as possible.

The isolation factor can be used to achieve a precise alignment of magnetic field to NV axis. According to Eq. (5), the off-axis field parallel to the test field $B_x$ should be minimized for a higher isolation factor. To verify both $B_x$ and $B_y$ are minimized, one needs an additional test field in different orientation. Thus, with the two test fields in different orientations, the peak intensity at the frequency of the test field in the NV thermometer will indicate the degree of the field alignment. Several off-axis field dependent phenomena have been reported, such as nuclear spin polarization at excited-state level anticrossing \cite{Jacques2009} and the decrease in photoluminescence intensity \cite{Tetienne2012}. Both may be used for the alignment of magnetic field, too. However, the former occurs only at a specific field near 50 mT, and the latter requires of a relatively strong field higher than 5 $\sim$ 50 mT. Equation (5) is considered more valid at weaker magnetic field, as it relies on the 2nd order approximation. Thus, the isolation factor can be a suitable indicator, exhibiting the degree of the alignment of a magnetic field of a few mT.

Of the four crystal axes in diamond, NV spins along the [111] orientation are exploited in the present work. As previously demonstrated, one can overlap all the four NV orientations by applying an external magnetic field along the [100] orientation to maximize ODMR contrast leading to enhanced sensitivity \cite{Barry2016}. NV spins which are not aligned to the external magnetic field have an issue of asymmetry in the NV resonance frequency shift due to external magnetic field variations, $\partial f / \partial B$; the ODMR peak at the lower mw frequency are less sensitive to magnetic fields than that at the higher mw frequency\cite{SM}. This asymmetry stems from the non linear response to the off-axis magnetic field in the presence of the zero-field splitting.  The discrepancy in $\partial f / \partial B$ will deteriorate the isolation factor as the first order terms ($\Delta B_z$) in Eq. (4) can not be cancelled with each other in the thermometry signal $S_T (t)$.

The shot-noise limits obtained from Eq. (1) and (2) are 13.5 pT/$\sqrt{\mathrm{Hz}}$ for the NV magnetometer and 5 $\mu$K/$\sqrt{\mathrm{Hz}}$ for the NV thermometer, values that are nearly a factor of five smaller than the experimental results. In Fig. 4 (a), the noise floor for magnetically-insensitive operation, when MW1 and MW2 are off, is about 41 pT/$\sqrt{\mathrm{Hz}}$, which is higher than the theoretical expectation. This discrepancy stems from incomplete cancellation of noise from the pumping laser. Additional noise, which appears when MW1 and MW2 are on, can be attributed to amplified microwaves; we found a higher microwave power increases the noise floor. In the present study, the spot of the pumping laser has a diameter of about 0.5 mm. Since the thickness of the diamond is 0.3 mm, the effective NV ensemble sensing volume is approximately 6 * 10$^7$ $\mu$m$^3$, and the volume-normalized sensitivity is 542 nT$\cdot$ $\mu$m$^{3/2}$/$\sqrt{\mathrm{Hz}}$. We noticed that increasing the spot size deteriorates the volume-normalized sensitivity significantly, consistent with there being spatial inhomogeneity across the NV ensemble of key parameters such as local strain, P1 and NV concentrations, external magnetic field, and mw power, etc. We note also that for the current set-up, absorption of the pump laser as well as NV fluorescence by P1 and NV centers in the diamond prevents even larger increase of the effective sensing volume.

\section{Conclusion}
In this work, we demonstrate a real-time multiplexed sensing of two physical quantities, magnetic field and temperature, using an ensemble of NV spins in diamond. Sensitivities as 70 pT/$\sqrt{\mathrm{Hz}}$ and 25 $\mu$K/$\sqrt{\mathrm{Hz}}$ are achieved for magnetic field and temperature, respectively, with good isolation of about 34 dB in temperature against magnetic field during simultaneous measurements. We developed a model of experimental parameters and their effect on NV measurements to help characterize and improve the isolation factor. The need for eliminating NV signal drifts caused by ambient thermal variation motivated this study. Such variations can come from many sources, e.g., thermal radiation, conduction through physical contacts, and convection of surrounding air flows. Thus, one must isolate thermal effects for reliable operation of an NV diamond magnetometer, and vice versa for an NV thermometer. Dual frequency driving makes simultaneous measurement of both quantities feasible; and its implementation is straightforward. Our method will extend NV diamond sensing application into a wider variety of fields. This may include operando \cite{Huang2020} or in-vitro \cite{Choi2020} monitoring, in which a chemical reaction occurs actively and brings charge currents with heat production. Additionally, a source of magnetic field can be highly temperature dependent, as is the case for magnetic nanoparticles having Curie temperature near room temperature \cite{Wang2018}. In such conditions, multiplexed NV diamond sensors will provide more accurate analysis by separating the effects of magnetic field and temperature.

\section*{Acknowledgement}
\label{Acknowledgement}
This research was supported by a grant (GP2021-0010) from Korea Research Institute of Standards and Science, and Institute of Information and communications Technology Planning $\&$ Evaluation (IITP) grant funded by the Korea government (MSIT) (No.2019-000296). Efforts of UMD QTC personnel (C.A.H., M.J.T., R.L.W.) was supported by the U.S. Army Research Laboratory MAQP program under Contract No. W911NF-19-2-0181.

\appendix

\section{Experimental methods}
\subsection{NV Diamond}
NV Diamond
The diamond is a general grade HPHT crystal (Element Six). The dimensions of the diamond were 3 * 3 * 0.3 mm$^3$ and the initial N$_S^0$ concentration was approximately 30 ppm. To create an NV- ensemble, the diamond was electron-irradiated (1 MeV and 1 * 10$^{19}$ e/cm$^2$) and annealed in vacuum at 950 $^{\circ}$C for 4 hours. The NV- concentration is 0.5 ppm (conversion efficiency ~ 1.7 \%) and remnant N$_S^0$ 1.3 ppm  \cite{SM, Howarth2003, Acosta2009}.

\subsection{Optical system}
The 532 nm pump laser (Millennia eV, 10 W) has an output power of 600 mW. The pump laser path is split by using a polarizing beam splitter and a half wave plate. The split path enters into the reference photodiode (Hamamatsu, S1337), monitoring the pump laser power. The fluorescence from the NV ensemble is collected by the signal photodiode (Hamamatsu, S1337). The photo-induced current is typically 11 mA. The reference and signal photodiodes are interconnected through a home-made balanced circuit, of which the output voltage is proportional to the difference between the photo-induced currents out of the reference and the signal photodiodes. Such balanced detection efficiently cancels the common noise from the pump laser. The power of the split pump laser is adjusted to balance the photo-induced currents on both photodiodes with 400 mW of green pump exciting the NV ensemble. The pump laser is focused onto the NV ensemble through a lens (f = 100 mm) with a typical spot size of 0.5 mm; and its optical polarization is adjusted to maximize the ODMR contrast of NV spins along the [111] orientation.
The half-ball lens is commercially available (Edmund) and made of a high refractive index material (S-LAH79, n = 1.987 at 700 nm). The elliptic reflector is manufactured via a high-precision machining (Nanoform L 1000) of an aluminum block. The surface roughness is typically less than 5 nm. The half-ball lens is glued to the NV diamond.

\subsection{Electronic system}
Two microwave sources (SG380) receive reference signals for the frequency modulation of MW1 and MW2. The reference signals are generated by a 2CH function generator (Agilent 33522A). For multiplexed sensing, the frequencies of the references $f_1^R$ and $f_2^R$ are set to 5 kHz and 7 kHz, respectively. A sinusoidal waveform was used for the reference signals, and the modulation depth was 0.55 MHz for both $f_1^R$ and $f_2^R$.
Frequency-modulated signals MW1 and MW2 are combined and put into a power amplifier (Minicircuit, 16 W). The output powers of MW1 and MW2 are typically 30 mW. Through a micro-strip line on PCB, the amplified microwave signal is guided to an $\Omega$-shaped antenna with a diameter of 7 mm and terminated with a 50 Ohm load. The antenna is mounted below a sapphire glass plate, upon which the NV diamond is positioned inside the elliptic reflector.
A permanent magnet (5 cm * 5 cm * 2.5 cm) produces an external magnetic field of about 1.6 mT at the position of the pump laser spot on the NV diamond. The permanent magnet is positioned on a combination of vertical and rotational stages, which allows careful alignment of the magnetic field along the [111] orientation. A Helmholtz coil, of diameter 20 cm, is used to produce test magnetic fields at 10 Hz. With a fluxgate sensor, the test magnetic field magnitude (RMS) is calibrated.
The sensitivity and time-constant of the two lock-in amplifiers (SR860) are set to 2 mV and 1 ms, respectively. A three-channel data-acquisition device is used for digitizing the NV fluorescence time trace signals with a device (NI-4461) of 24-bit resolution and a sampling rate of 1 kS/s. The first and second channels are connected to the outputs of LIA1 and LIA2, respectively. The third channel measures the voltages across the PT100 temperature sensor via 4-probe method. An external source applies 10 mA current through the PT100 sensor. The resistance of the PT100 sensor was calibrated as a function of temperature prior to NV measurements.

\subsection{Data processing}
For MW1 and MW2 used for dual frequency driving, power and modulation depths are adjusted not only for optimizing NV sensitivity but also for balancing to have nearly the same signal in the two detection channels in the presence of the test signal. For a high isolation factor during multiplexed sensing, such balancing is required but imperfect, due to the difficulty in matching two large signals with a high degree of precision. Thus, we perform an additional process; the digitized time trace signals ($S_1$ and $S_2$) from LIA1 and LIA2 are summed ($S_T$) and subtracted ($S_B$), following the equations below,
\begin{eqnarray}
S_T & = & \sqrt{2} \left[ \cos(\frac{\pi}{4} + \epsilon) S_1 + \sin(\frac{\pi}{4} + \epsilon) S_2 \right] \nonumber \\
S_B & = & \sqrt{2} \left[ \cos(\frac{\pi}{4} + \epsilon) S_1 - \sin(\frac{\pi}{4} + \epsilon) S_2 \right].
\end{eqnarray}
The phase $\epsilon$ is digitally tuned with a precision of 0.01$^{\circ}$ to minimize the intensity of the test signal on $S_T$. Typically $\epsilon$ varies within $\pm$1.5$^{\circ}$.

\bibliography{bibliography}

\end{document}